\documentclass[reprint, superscriptaddress, amsmath, amssymb, aps, prb]{revtex4-2}
\usepackage{graphicx}
\usepackage{dcolumn}
\usepackage{bm}
\usepackage{placeins}
\usepackage{appendix}

\usepackage{verbatim}
\usepackage[usenames,dvipsnames]{xcolor}
 \usepackage{amsmath}
 \usepackage{amsfonts}
 \usepackage{amssymb}
 \usepackage[colorlinks=true,citecolor=blue,linkcolor=red]{hyperref}

 \usepackage{bbold}     
 \usepackage[makeroom]{cancel}  
 \usepackage{multirow}    
 \usepackage[normalem]{ulem}        
 \usepackage{array}
 \usepackage{pdfpages}
\makeatletter
 \AtBeginDocument{\let\LS@rot\@undefined}
 \makeatother

\begin{document}

\title{Observation of Aharonov-Bohm effect in PbTe nanowire networks}

\author{Zuhan Geng}
\email{equal contribution}
\affiliation{State Key Laboratory of Low Dimensional Quantum Physics, Department of Physics, Tsinghua University, Beijing 100084, China}

\author{Zitong Zhang}
\email{equal contribution}
\affiliation{State Key Laboratory of Low Dimensional Quantum Physics, Department of Physics, Tsinghua University, Beijing 100084, China}

\author{Fangting Chen}
\email{equal contribution}
\affiliation{State Key Laboratory of Low Dimensional Quantum Physics, Department of Physics, Tsinghua University, Beijing 100084, China}

\author{Shuai Yang}
\affiliation{Beijing Academy of Quantum Information Sciences, 100193 Beijing, China}

\author{Yuying Jiang}
\affiliation{State Key Laboratory of Low Dimensional Quantum Physics, Department of Physics, Tsinghua University, Beijing 100084, China}

\author{Yichun Gao}
\affiliation{State Key Laboratory of Low Dimensional Quantum Physics, Department of Physics, Tsinghua University, Beijing 100084, China}

\author{Bingbing Tong}
\affiliation{Beijing Academy of Quantum Information Sciences, 100193 Beijing, China}

\author{Wenyu Song}
\affiliation{State Key Laboratory of Low Dimensional Quantum Physics, Department of Physics, Tsinghua University, Beijing 100084, China}

\author{Wentao Miao}
\affiliation{State Key Laboratory of Low Dimensional Quantum Physics, Department of Physics, Tsinghua University, Beijing 100084, China}

\author{Ruidong Li}
\affiliation{State Key Laboratory of Low Dimensional Quantum Physics, Department of Physics, Tsinghua University, Beijing 100084, China}

\author{Yuhao Wang}
\affiliation{State Key Laboratory of Low Dimensional Quantum Physics, Department of Physics, Tsinghua University, Beijing 100084, China}

\author{Qinghua Zhang}
\affiliation{Institute of Physics, Chinese Academy of Sciences, Beijing 100190, China}

\author{Fanqi Meng}
\affiliation{Institute of Physics, Chinese Academy of Sciences, Beijing 100190, China}

\author{Lin Gu}
\affiliation{Institute of Physics, Chinese Academy of Sciences, Beijing 100190, China}

\author{Kejing Zhu}
\affiliation{Beijing Academy of Quantum Information Sciences, 100193 Beijing, China}

\author{Yunyi Zang}
\affiliation{Beijing Academy of Quantum Information Sciences, 100193 Beijing, China}

\author{Lin Li}
\affiliation{Beijing Academy of Quantum Information Sciences, 100193 Beijing, China}

\author{Runan Shang}
\affiliation{Beijing Academy of Quantum Information Sciences, 100193 Beijing, China}

\author{Xiao Feng}
\affiliation{State Key Laboratory of Low Dimensional Quantum Physics, Department of Physics, Tsinghua University, Beijing 100084, China}
\affiliation{Beijing Academy of Quantum Information Sciences, 100193 Beijing, China}
\affiliation{Frontier Science Center for Quantum Information, 100084 Beijing, China}

\author{Qi-Kun Xue}
\affiliation{State Key Laboratory of Low Dimensional Quantum Physics, Department of Physics, Tsinghua University, Beijing 100084, China}
\affiliation{Beijing Academy of Quantum Information Sciences, 100193 Beijing, China}
\affiliation{Frontier Science Center for Quantum Information, 100084 Beijing, China}
\affiliation{Southern University of Science and Technology, Shenzhen 518055, China}

\author{Ke He}
\email{kehe@tsinghua.edu.cn}
\affiliation{State Key Laboratory of Low Dimensional Quantum Physics, Department of Physics, Tsinghua University, Beijing 100084, China}
\affiliation{Beijing Academy of Quantum Information Sciences, 100193 Beijing, China}
\affiliation{Frontier Science Center for Quantum Information, 100084 Beijing, China}

\author{Hao Zhang}
\email{hzquantum@mail.tsinghua.edu.cn}
\affiliation{State Key Laboratory of Low Dimensional Quantum Physics, Department of Physics, Tsinghua University, Beijing 100084, China}
\affiliation{Beijing Academy of Quantum Information Sciences, 100193 Beijing, China}
\affiliation{Frontier Science Center for Quantum Information, 100084 Beijing, China}


\begin{abstract}

We report phase coherent electron transport in PbTe nanowire networks with a loop geometry. Magneto-conductance shows Aharonov-Bohm (AB) oscillations with periods of $h/e$ and $h/2e$ in flux. The amplitude of $h/2e$ oscillations is enhanced near zero magnetic field, possibly due to interference between time-reversal paths. Temperature dependence of the AB amplitudes suggests a phase coherence length $\sim$ 8 - 12 $\mu$m at 50 mK. This length scale is larger than the typical geometry of PbTe-based hybrid semiconductor-superconductor nanowire devices.

\end{abstract}

\maketitle

\section{Introduction}

Topological quantum computing relies on braiding of Majorana zero modes (MZMs) to realize various quantum gate operations \cite{Kitaev_2003, Nayak}. A major theoretical proposal is the so-called measurement-based braiding where the topological qubit can be readout from interference between a MZM-involved path (electron `teleportation') and a normal (topologically trivial) path \cite{Fu_2010,Plugge_2017,Fu_2016,Scalable_2017}. Therefore, phase coherent transport is a crucial ingredient in this roadmap. One-dimensional semiconductor-superconductor hybrid nanowire is a promising MZM candidate \cite{Lutchyn2010, Oreg2010, Prada2020,NextSteps} where the phase coherent transport can be revealed as Aharonov-Bohm (AB) oscillations in nanowire networks with a loop geometry. Among all 1D candidates, InSb and InAs nanowires are the mostly studied  MZM material systems \cite{Mourik, Deng2016, Albrecht, Gul2018, Zhang2021,Song2021}, with AB effect demonstrated \cite{Gazibegovic2017,CPH_SAG_2018,SB_SAG,Delft_SAG_InSb,Roy_SAG}. Recently, PbTe \cite{Poland_1993, Poland_1999, Poland_2004, Poland_PRB_2005, Poland_2006, Poland_2006_QPC} based nanowires \cite{Erik_PbTe, Jiangyuying} have been proposed as a new MZM candidate \cite{CaoZhanPbTe}. In this paper, we report the observation of AB effect in PbTe nanowire networks. 

\begin{figure}[ht]
\includegraphics[width=\columnwidth]{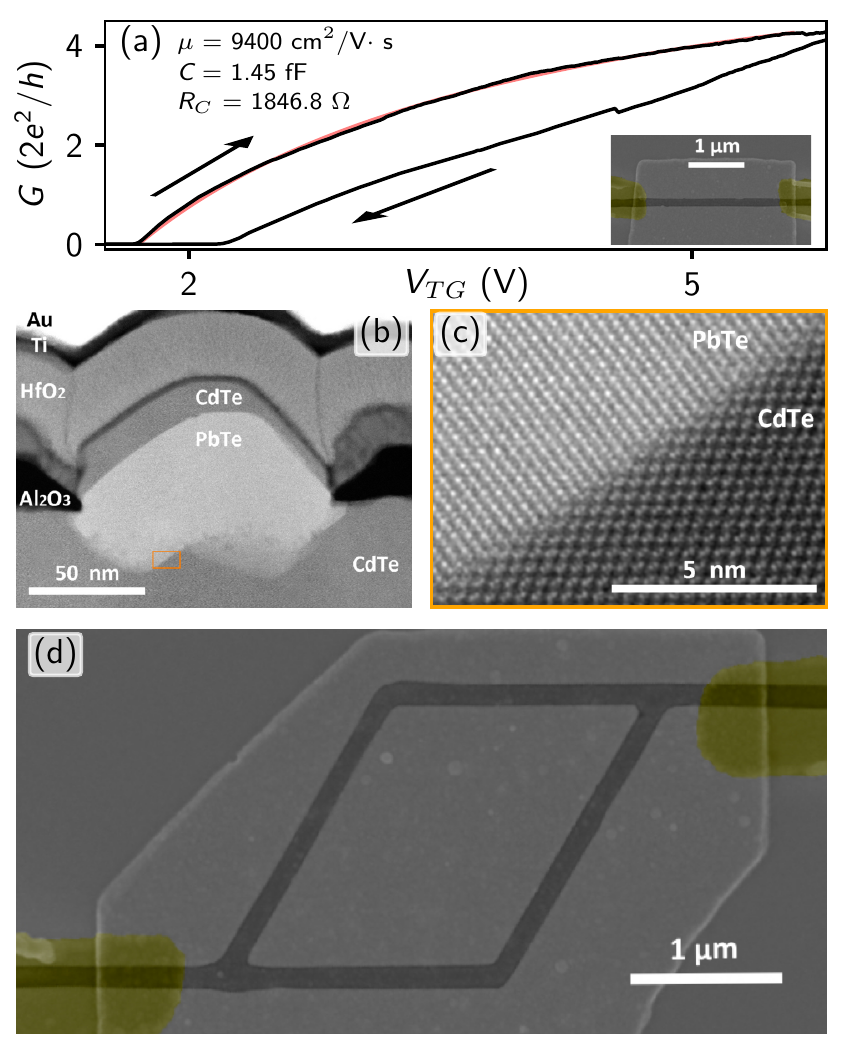}
\centering
\caption{(a) Pinch-off of a PbTe nanowire (inset). Red line, mobility fit. (b) Device cross section. (c) TEM of the PbTe-CdTe(111) interface. (d) SEM of Device A (nanowire loop). }
\label{fig1}
\end{figure}

\section{Device basics}

\begin{figure*}[tb]
\includegraphics[width=\textwidth]{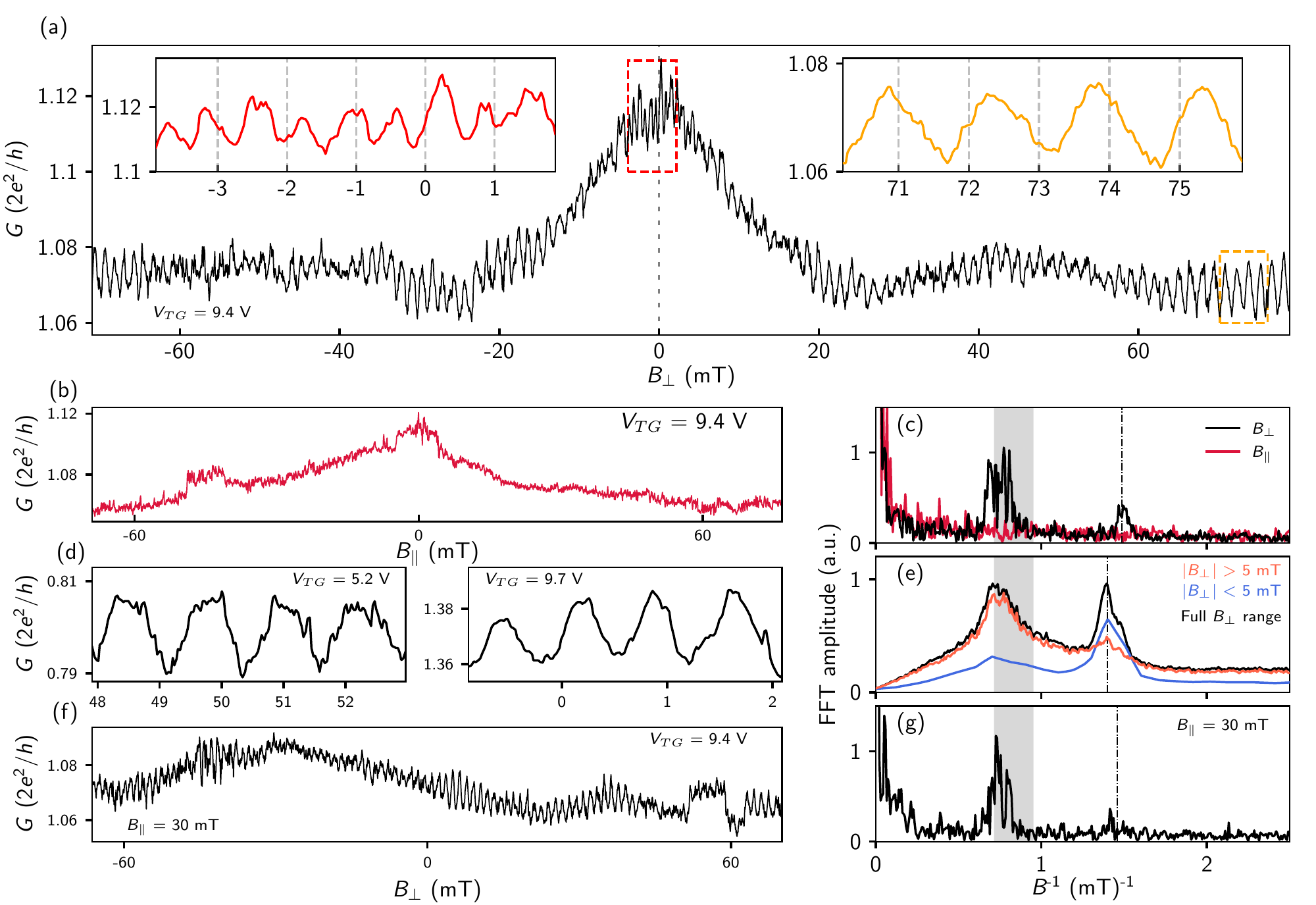}
\centering
\caption{AB oscillations in Device A. (a) Magneto-conductance, $B$ perpendicular to the substrate. Insets, magnifications of the regions of $h/2e$ and $h/e$ oscillations. (b) Magneto-conductance, $B$ parallel to the substrate. $B$ is offset by 4 mT and 2 mT for (a) and (b), based on the symmetry of the WAL.  (c) FFT of (a) (black) and (b) (red). (d) Two more AB curves, $B$ perpendicular to the substrate. (e) Black: ensemble averaged FFT, after subtracting a conductance background (smooth window 1.42 mT $\sim h/e$ period). Blue: ensemble averaged FFT by limiting the $B$-range below 5 mT. Red: averaged FFT outside this range. (f) AB at a fixed in-plane $B$ of 30 mT. (g) FFT of (f). In (c), (e) and (g), the grey backgrounds are the geometric bounds estimated from the device SEM. The vertical dashed line is the expected position for $h/2e$. Fridge temperature $\sim$ 20 mK.}
\label{fig2}
\end{figure*}

\begin{figure*}[tb]
\includegraphics[width=\textwidth]{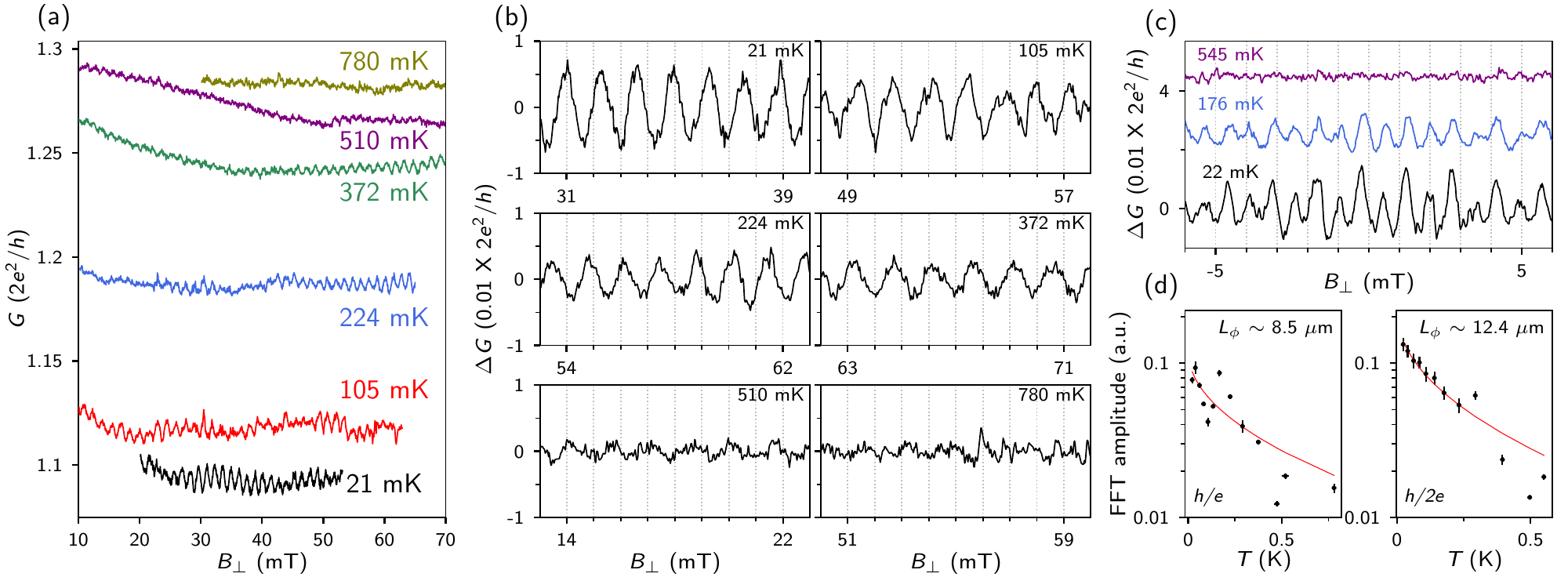}
\centering
\caption{Temperature ($T$) dependence of AB oscillations in Device A. (a) $h/e$ oscillations at several $T$s. $V_{TG}$ = 9.4 V. (b) AB after subtracting a background (smooth window 1.42 mT). (c) $h/2e$ oscillations (background subtracted) at three $T$s. Vertical offsets are 0.025 and 0.045 ($\times 2e^2/h$) for clarity. $V_{TG}$ = 9.8 V. (d) $T$-dependence of the $h/e$ (left) and $h/2e$ (right) FFT amplitudes for all measured $T$s. Red lines are fittings to extract the phase coherence length. }
\label{fig3}
\end{figure*}

Fig. 1a (inset) shows the scanning electron micrograph (SEM) of a PbTe nanowire, selectively grown on a CdTe substrate using molecular beam epitaxy (MBE). The nanowire is contacted by two normal metal leads (yellow, Ti/Au) and then covered by a layer of HfO$_2$ dielectric with a Ti/Au top gate. The selective area growth (SAG) of PbTe is achieved by covering the CdTe substrate with an Al$_2$O$_3$ dielectric mask followed by wet etched trenches. Growth details can be found in our recent work \cite{Jiangyuying} where a CdTe(001) substrate was used. In this work, we choose a different substrate orientation, CdTe(111), for the PbTe growth. Valley-degeneracy breaking along this (111) direction, which is crucial in the realization of MZMs, has been experimentally demonstrated \cite{Poland_PRB_2005,Poland_2006_QPC} and theoretically verified \cite{CaoZhanPbTe}. 

Mobility fit (for the upward sweeping direction) of this field effect device (Fig. 1a) suggests a mobility $\sim$ 0.94 $\times 10^4$ cm$^2$/Vs, with the capacitance ($C$) estimated using a finite element model based on the device cross section (Fig. 1b). Clearly, the cross section has an irregular shape with rough interfaces, possibly due to the Ar treatment of the CdTe substrate before the PbTe growth. This interface can be a major source of disorder \cite{GoodBadUgly,DasSarma_disorder_2021,Tudor2021Disorder, DasSarma_estimate,PRApplied_DasSarma} and future growth optimization for flat interface may lead to higher mobility. Nevertheless, magnification of the interface (Fig. 1c) can still resolve matched lattices between PbTe and CdTe. Fig. 1d shows the SEM of a PbTe network device (Device A), grown and fabricated (contact, dielectric and gate) together with the mobility device on the same substrate chip. Transmission electron microscopy (TEM) analysis of Device A can be found in Supplementary (Sfig. 1).

\section{AB oscillations in Device A}

Fig. 2a shows the magneto-conductance of Device A, measured in a dilution refrigerator with a base temperature $\sim$ 20 mK. The differential conductance $G$ was measured using a lock-in amplifier in a two terminal circuit set-up. The bias voltage was fixed at zero throughout the measurement. The magnetic field ($B$) was oriented perpendicular to the substrate for Fig. 2a. The overall conductance shows a peak near $B$ = 0 T, suggesting the existence of weak anti-localization (WAL). On top of the background, periodic AB oscillations are resolved with periods of $\Delta B \sim$ 1.4 mT (orange inset) and $\sim$ 0.7 mT (red inset), corresponding to flux periods of $h/e$ (first harmonic) and $h/2e$ (second harmonic), respectively. Here, $h$ is the Plank constant and $e$ the electron charge. We further convert this period in $\Delta B$ into an effective loop area $A \sim$ 2.95 $\mu$m$^2$ based on the formula of $\Delta B\times A = h/e$. This extracted area is close to the area defined by the inner surface of the nanowire loop ($\sim$ 2.9 $\mu$m$^2$), suggesting that the electron wave-function is mainly distributed near the inner surface.

For a control test, we also apply $B$ parallel to the substrate and observe a WAL peak without AB oscillations (Fig. 2b). In Fig. 2c we plot the spectrum of Fast Fourier Transform (FFT) where the $h/e$ and $h/2e$ oscillations in Fig. 2a are revealed as two FFT peaks (black curve). By contrast, the FFT of Fig. 2b shows no such peaks (the red curve in Fig. 2c). Boundaries of the shaded area refer to the expected $h/e$ AB periods estimated based on the area encircled by the inner and outer surfaces of the nanowire loop. The dashed line is the expected peak position of the second harmonic ($h/2e$), calculated by multiplying the center of the first harmonic peak by a factor of two.

We further show the magneto-conductance of Device A at different gate voltages, see supplementary (Sfig. 2). The AB amplitude varies for different gate voltages, as well as for the same gate voltage but repeated measurements, possibly due to device instabilities. In some gate settings, we observe no AB oscillations, possibility due to the pinch-off of one of the two AB `arms' or other unknown dephasing mechanisms. Fig. 2d shows examples for $h/e$ and $h/2e$ oscillations with amplitudes reaching $\sim$ 0.015 and 0.02, in unit of $2e^2/h$. In Fig. 2e we show the ensemble averaged FFT for all (in total 182 sets) measured $B$-sweeping curves (the black line), where the first and second harmonic peaks are clearly visible. The second harmonic peak has a similar height as the first one, suggesting a significant $h/2e$-component. The $h/2e$ oscillations in Fig. 2a are mostly prominent near $B$ = 0 T, possibly due to the time-reversal paths. To confirm this hypothesis, we perform the ensemble averaged FFT for $B$ near 0 T (blue curve in Fig. 2e) which indeed resolves a more prominent $h/2e$ peak. By contrast, the averaged FFT for higher $B$ resolves a much smaller $h/2e$ peak (red curve in Fig. 2e). In addition, in Fig. 2f with a small fixed in-plane magnetic field of 30 mT, the out-of-plane magneto-conductance reveals mainly the $h/e$ oscillations even for out-of-plane $B$ near 0 T. The corresponding FFT (Fig. 2g) also confirms that the $h/2e$ peak is hardly visible.

\section{Temperature Dependence}

We now study the temperature dependence of the AB oscillations. Fig. 3a shows the $h/e$ oscillations with temperatures vary from 21 mK to 780 mK (only six curves are shown for clarity). The $T$-evolution of AB amplitudes is more visible after subtracting an overall background (by smoothing the curve), as shown in Fig. 3b which focuses on the $B$-ranges with large oscillations. The oscillation amplitude is reduced by roughly half at $T \sim$ 200 - 300 mK, and almost vanishes at $T \sim$ 780 mK. Fig. 3c shows the $h/2e$ oscillation (background subtracted) at three typical $T$s, resolving similar trend. 

\begin{figure}[htb]
\includegraphics[width=\columnwidth]{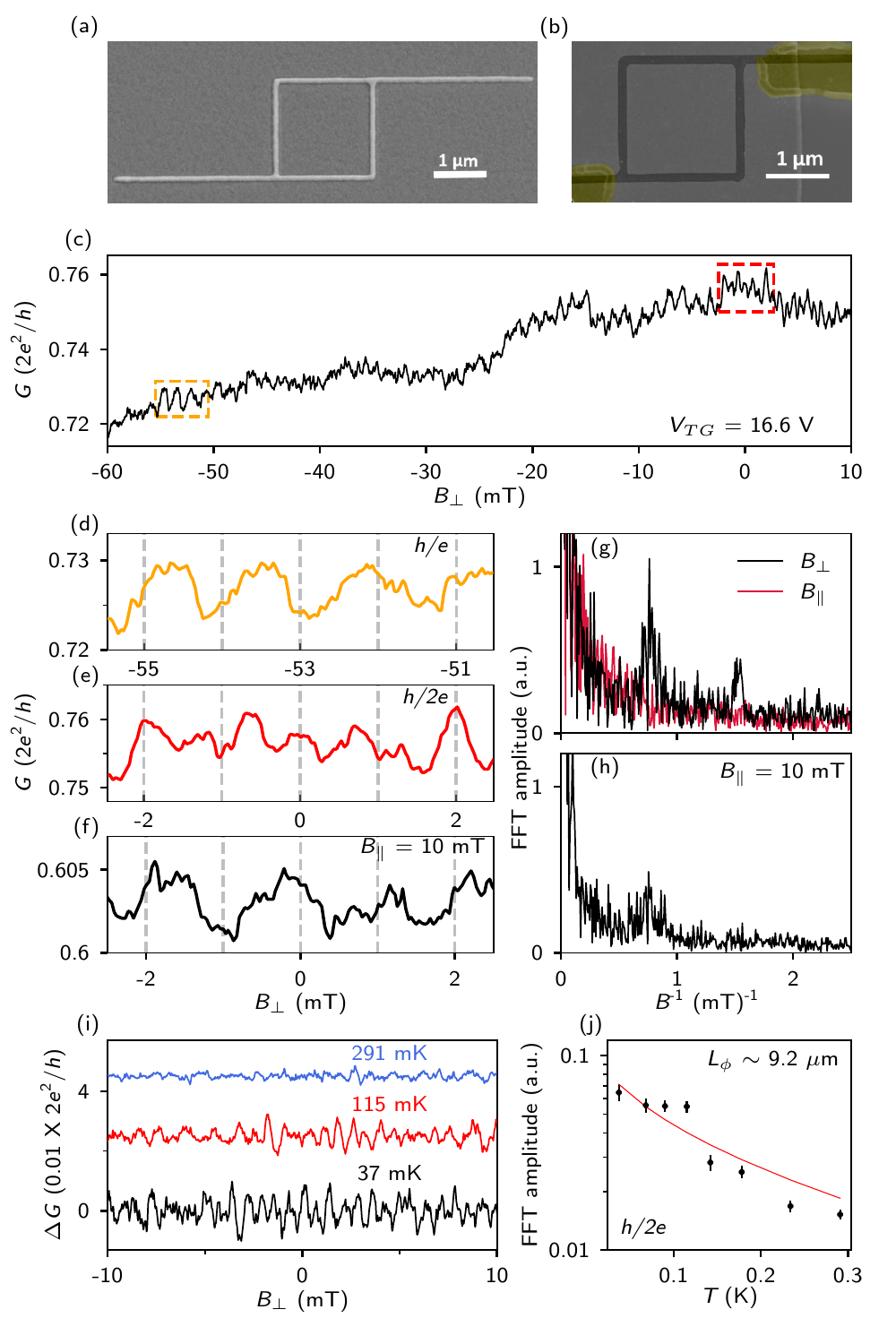}
\centering
\caption{AB effect in Device B. (a-b) SEM before (a) and after (b) the fabrication of contacts (false colored yellow), dielectric and top gate. (c) Magneto-conductance, $B$ perpendicular to the substrate. A narrower $B$-range is shown for clarity. Fridge temperature $\sim$ 40 mK. (d-e) Magnifications of (c) (see dashed boxes), corresponding to $h/e$ and $h/2e$ oscillations. (f) Magnification of a $B$-scan with a fixed in-plane $B$ of 10 mT. $V_{TG}$ = 16.6 V. $B$-offset is  5 mT for (c-f).  (g) Black: FFT spectrum of (c). Red: FFT of a $B$-scan (not shown) at the same gate voltage, $B$ parallel to the substrate, serving as a control test. (h) FFT of (f) (the whole $B$-range). (i) $h/2e$ oscillations at three different $T$s with background conductance subtracted. Vertical offsets are 0.025 and 0.045 $\times 2e^2/h$ for clarity. Background is smoothed over a window of 1.34 mT, close to the $h/e$ period estimated based on the inner surface loop area. (j) $T$-dependence of the $h/2e$ FFT amplitudes. The fitting (red line) suggests a phase coherence length $\sim$ 9.2 $\mu$m at 50 mK.}
\label{fig4}
\end{figure}

To quantify the $h/e$ and $h/2e$ oscillation amplitudes, we perform FFT and estimate the amplitude by the `area' underneath the corresponding FFT peak. This area is calculated by numerical integration, see Sfig. 3 for detailed information. Fig. 3d shows $h/e$ (left) and $h/2e$ (right) amplitudes decaying with increasing $T$ as a general trend. The fluctuations is possibly due to instabilities within the mesoscopic environment. To fit this decay, we adopt a diffusive transport model \cite{Albert_AB, Delft_SAG_InSb}, where the AB amplitude is assumed to be $\propto e^{-L/L_{\varphi}}$ and $L_{\varphi}$ assumed to be $\propto 1/\sqrt{T}$. $L_{\varphi}$ is the phase coherence length and $L$ is one half of the loop circumference for $h/e$ and the full circumference for $h/2e$ oscillations. The fitting result (red lines in Fig. 3d) suggests a phase coherence length $L_\varphi$ $\sim$ 8 - 12 $\mu$m at a temperature of 50 mK. We note this $L_\varphi$ value is only a rough estimation which may vary for different models. The extracted length scale (a few micron) is consistent with the observation of $h/2e$ oscillations: electrons remain phase coherent after circulating the whole loop which is $\sim$ 8 $\mu$m.

\section{AB oscillations in another Device}

Finally, we show AB oscillations observed in a second device (Device B). The loop geometry has a square shape instead of a diamond as shown in Fig. 4a-b. The magneto-conductance (Fig. 4c) resolves AB oscillations with $h/2e$ period near zero $B$ and $h/e$ period at higher $B$, see Fig. 4d-e for the magnifications with corresponding colors. The AB periods for $h/e$ and $h/2e$ oscillations are $\sim$ 1.25 and 0.66 mT, respectively. This period in $B$ can be converted into an effective loop area of $\sim$ 3.3 $\mu$m$^2$. Based on the device SEM, we can independently estimate the loop area to be $\sim$ 3 and 4 $\mu$m$^2$ for the inner and outer surfaces, respectively. Comparison to the evaluated loop area again suggests that the electrons are mainly located near the inner surface of the loop.

Fig. 4g shows the FFT of Fig. 4c, resolving the two harmonic peaks (black curve). As a comparison, FFT of a $B$-scan curve (not shown), measured under the same gate voltage but with $B$ parallel to the substrate, resolves no such peaks (red curve). The $h/2e$ oscillations are also enhanced near zero $B$, similar to the case in Device A. This $h/2e$ component can be reduced by a fixed in-plane $B$ of 10 mT, as shown in Fig. 4f. The corresponding FFT (for the whole range) also only resolves the $h/e$ peak (Fig. 4h).

Due to the device instabilities, we have only performed temperature dependence for the $h/2e$ oscillations, see Fig. 4i for the oscillations at three different temperatures. In Fig. 4j, we plot the FFT amplitudes vs temperature, fit the diffusive transport model and extract the phase coherence length $L_{\phi}\sim 9$ $\mu$m at $T$ = 50 mK. This result is roughly consistent with the value of Device A. For more $B$-scans of Device B, see Sfig. 2.

\section{Summary}

To summarize, we have demonstrated phase coherent transport in PbTe nanowire networks by observing Aharonov-Bohm oscillations in magneto-conductance. Both $h/e$- and $h/2e$-period oscillations can be revealed. Temperature dependence of the AB amplitude suggests a phase coherence length $\sim$ 8 - 12 $\mu$m at $T$ = 50 mK. This length scale exceeds the dimension of future PbTe-based hybrid semiconductor-superconductor devices, fulfilling a necessary condition for the exploration of topological quantum information processing. Future works could be aiming at the optimization of the nanowire growth for more uniform cross-sections which may lead to higher mobility and longer phase coherence length.

\section{Acknowledgment} 

We thank Gu Zhang and Zhan Cao for valuable discussions. This work is supported by Tsinghua University Initiative Scientific Research Program, National Natural Science Foundation of China (92065206, 51788104) and National Key Research and Development Program of China (2017YFA0303303). Raw data and processing codes within this paper are available at https://zenodo.org/record/5798424.

\bibliography{mybibfile}

\newpage

\onecolumngrid
\newpage


\section*{Supplementary material}

# Supplement Information: Observation of Aharonov-Bohm effect in PbTe nanowire networks

Zuhan Geng,[1, *] Zitong Zhang,[1, *] Fangting Chen,[1, *] Shuai Yang,[2] Yuying Jiang,[1] Yichun Gao,[1] Bingbing Tong,[2] Wenyu Song,[1] Wentao Miao,[1] Ruidong Li,[1] Yuhao Wang,[1] Qinghua Zhang,[3] Fanqi Meng,[3] Lin Gu,[3] Kejing Zhu,[2] Yunyi Zang,[2] Lin Li,[2] Runan Shang,[2] Xiao Feng,[1, 2, 4] Qi-Kun Xue,[1, 2, 4, 5] Ke He,[1, 2, 4, †] and Hao Zhang[1, 2, 4, ‡]

[1]*State Key Laboratory of Low Dimensional Quantum Physics,
Department of Physics, Tsinghua University, Beijing 100084, China*
[2]*Beijing Academy of Quantum Information Sciences, 100193 Beijing, China*
[3]*Institute of Physics, Chinese Academy of Sciences, Beijing 100190, China*
[4]*Frontier Science Center for Quantum Information, 100084 Beijing, China*
[5]*Southern University of Science and Technology, Shenzhen 518055, China*

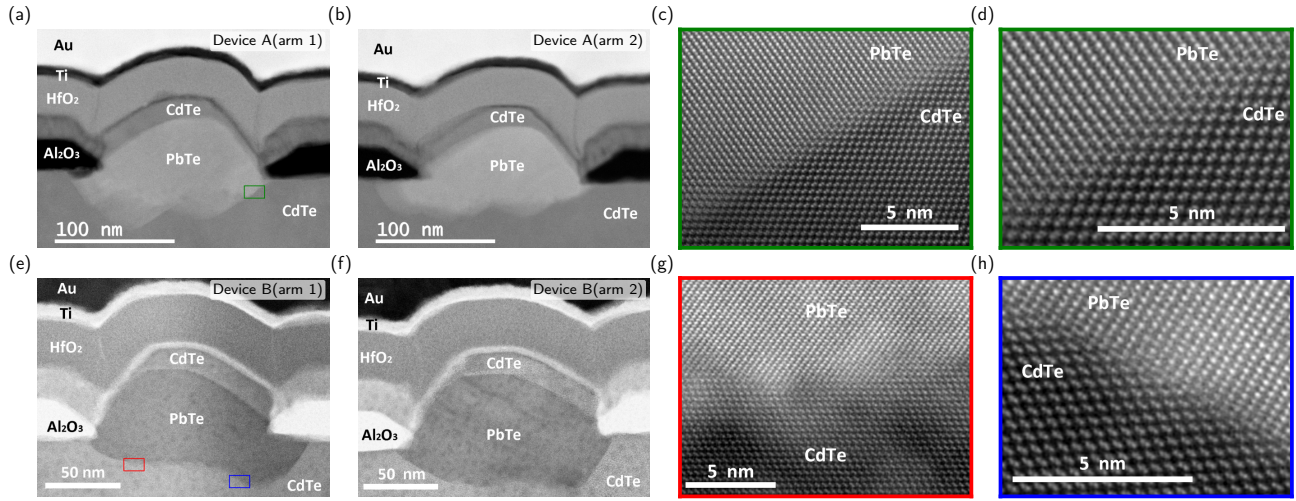

Supplement Fig. 1: TEM analysis of PbTe(111) AB nanowire loops. (a) and (b) are the cross section images of the two AB 'arms' of Device A. (c) is the magnification of (a) (indicated by the box). (d) is a magnification of (c). (e) and (f) are the cross section images of the two AB 'arms' of Device B. (g-h) are the magnifications of (e-f) (indicated by the boxes). The interface between CdTe and PbTe in (g) suggests an irregular and non-uniform CdTe substrate surface after the Ar treatment (before PbTe growth), which is also evident in (a-b) and (e-f). Future optimization for flat substrate surface may further decrease this disorder and lead to better transport properties (e.g. higher mobility and longer phase coherence length).

---

* equal contribution
† kehe@tsinghua.edu.cn
‡ hzquantum@mail.tsinghua.edu.cn

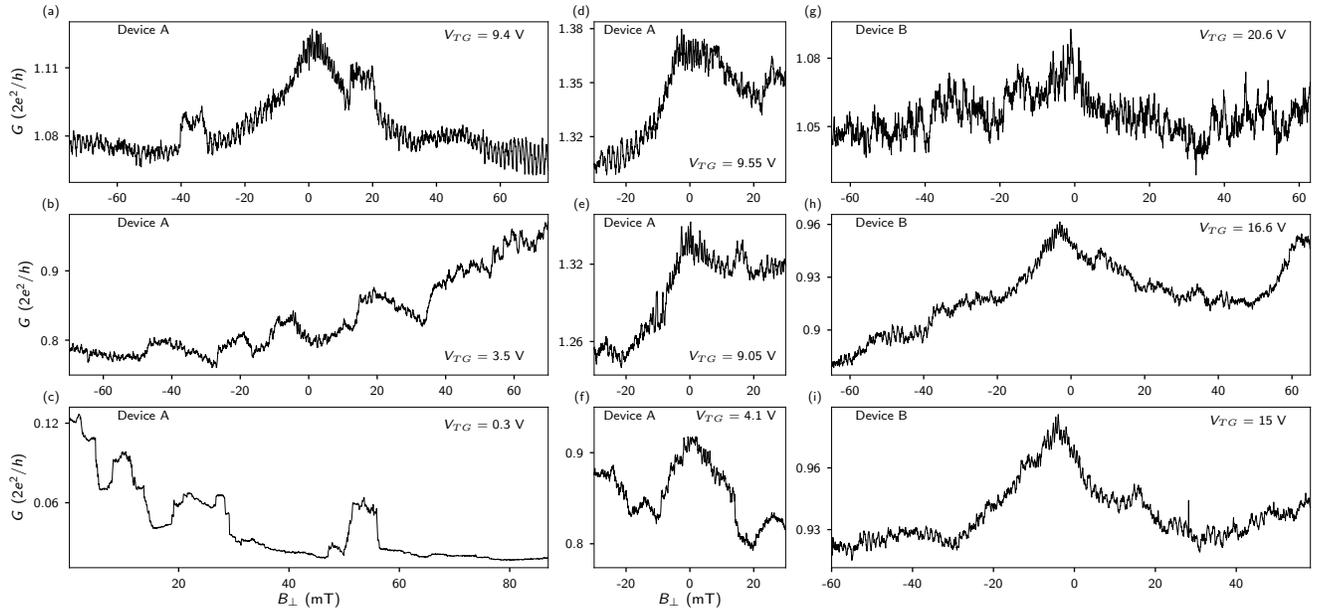

Supplement Fig. 2: Additional $B$-scans of Device A (a-f) and Device B (g-i), $V_{TG}$ labeled in each panel. $T \sim 20$ mK for Device A and 40 mK for Device B. At some gate settings, e.g. (c) and (g), no AB oscillation is observed. The WAL is also absent in e.g. (b), (d) and (g).

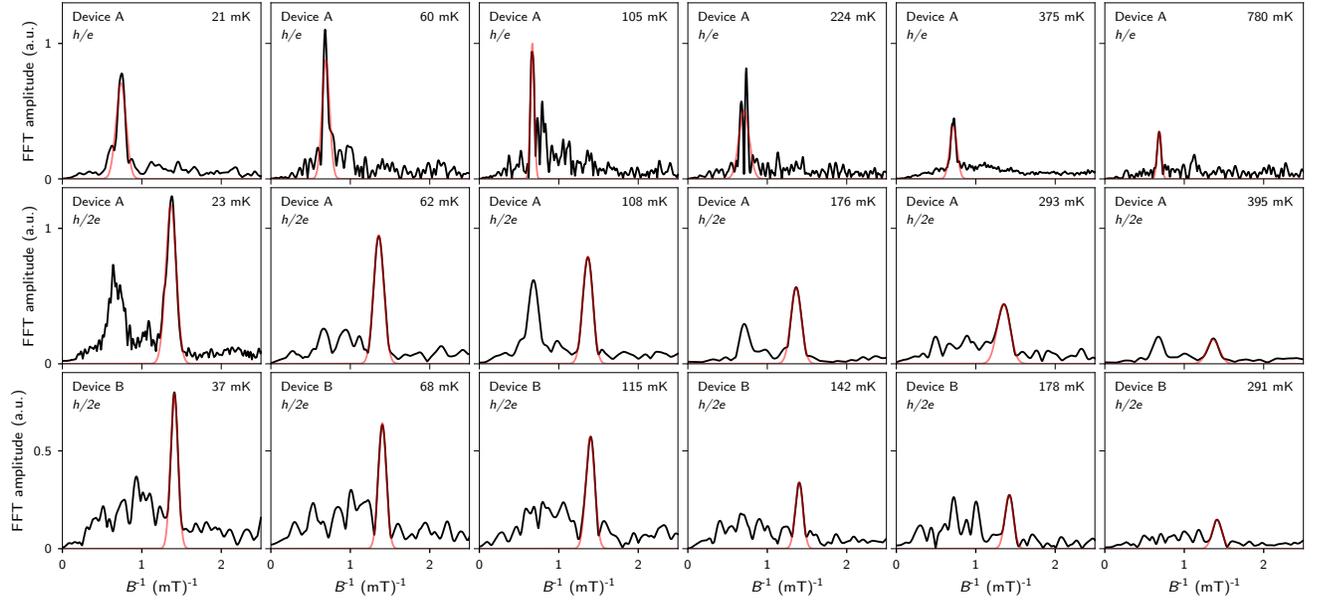

Supplement Fig. 3: Extraction of FFT amplitudes. The FFT amplitudes for $T$-dependence are estimated in the following. 1) The raw data of each $B$-scan curve is interpolated onto a uniformly spaced grid with 50 data points per mT. 2) A smoothed background is subtracted with a smooth window of 1.42 mT for device A and 1.34 mT for device B (close to the estimated periods of $h/e$ oscillations). 3) FFT is performed with a Hanning window. 4) For each $T$, we have measured several $B$-scan curves. The FFTs (absolute value) of these curves, corresponding to the same $T$, are averaged and shown in panels above as black lines (six temperatures for each series are shown for clarity). 5) A Gaussian line shape is fitted to the FFT spectrum near the harmonic peak (see red lines in each panel), where the standard deviation ($\sigma$) and the mean value ($\mu$) can be extracted. 6) The FFT spectrum is integrated between $\mu - \sigma$ and $\mu + \sigma$. The corresponding area within this range is used for the FFT amplitude. 7) The error bar of the FFT amplitude is estimated based on the integral difference between the interval $[\mu - \sigma, \mu + \sigma]$ to $[\mu - \sigma - \frac{1}{2B}, \mu + \sigma - \frac{1}{2B}]$ and $[\mu - \sigma + \frac{1}{2B}, \mu + \sigma + \frac{1}{2B}]$, where $B$ is the range of the corresponding $B$-scan.



\end{document}